\documentclass[10pt,conference]{IEEEtran}

\usepackage{cite}
\usepackage{amsmath,amssymb,amsfonts}
\usepackage{subfig}
\usepackage{algorithmic,algorithm2e}
\usepackage{graphicx}
\usepackage{textcomp}
\usepackage{xcolor}
\usepackage{color}
\usepackage{colortbl}
\usepackage{url}
\usepackage{listings}
\usepackage{booktabs}

\def\BibTeX{{\rm B\kern-.05em{\sc i\kern-.025em b}\kern-.08em
    T\kern-.1667em\lower.7ex\hbox{E}\kern-.125emX}}

\newtheorem{definition}{Definition}

\begin{document}

\title{Towards the Use of Slice-based Cohesion Metrics with Learning Analytics to Assess Programming Skills}

\author{\IEEEauthorblockN{Max Kesselbacher}
\IEEEauthorblockA{\textit{Department of Informatics Didactics} \\
\textit{University of Klagenfurt}\\
Klagenfurt, Austria \\
max.kesselbacher@aau.at}
\and
\IEEEauthorblockN{Andreas Bollin}
\IEEEauthorblockA{\textit{Department of Informatics Didactics} \\
\textit{University of Klagenfurt}\\
Klagenfurt, Austria \\
andreas.bollin@aau.at}
}

\maketitle

\begin{abstract}
In programming education, it makes a difference whether you are dealing with beginners or advanced students. As our future students will become even more tech-savvy, it is necessary to assess programming skills appropriately and quickly to protect them from boredom and optimally support the learning process. In this work, we advocate for the use of slice-based cohesion metrics to assess the process of program construction in a learning analytics setting. We argue that semantically related parts during program construction are an essential part of programming skills. Therefore, we propose using cohesion metrics on the level of variables to identify programmers' trains of thought based on the cohesion of semantically related parts during program construction.
\end{abstract}


\section{Introduction}

In his WiPSCE'16 keynote \cite[p.~14]{Lister2016}, Raymond Lister stressed that, in his (neo-Piagetian) view\footnote{In the neo-Piagetian view of Lister, learning to program occurs in four main stages of overlapping waves, with the ability to mentally execute, i.e. \textit{trace}, program code as a cornerstone of novice programmers' progression.}, quite some steps are necessary to master writing one's own first programs, and curricula and teachers are not adequately prepared for facilitating the learning process. The situation seems to continue at the university level, where high failure rates are observed in programming courses \cite{Bornat08}, or are at least compounded by faulty teaching strategies \cite{Luxton-Reilly16}.

Even if one is not a supporter of (neo-)Piagetian ideas, to develop the programming skills of our future software engineers in the best possible way, it is essential to understand the learning processes, to know more about the strategies experts follow when constructing their program, and to learn how they are mapping their trains of thought to running code. Early work by Weiser \cite{Weiser1982}, Burnstein et al.~\cite{burn98a}, Broad and Filer \cite{broad99} and also the application in the formal methods domain \cite{Bollin2013} already showed that programmers think in structures detectable by slices. This makes slice-based program analysis a promising avenue towards our aim, namely identifying code construction patterns and strategies that can then be used to support novice users in learning to program.

Slicing techniques are usually applied on finished programs or successive program versions stored in software version control systems. With the emergence of learning analytics (LA) \cite{Ihantola2015, Hundhausen2017}, there is a great potential to improve individual assessment and feedback, even during program construction. This strongly affects the next generation of software engineers: with heterogenous previous programming experiences and diversity in the possible tools for learning, they need to be supported individually; a prospect made possible with LA.

We argue that the granularity of current slice metrics does not fit our established aims: assessing the \textit{method-level} cohesion \cite{Ott1993} alone is too coarse for individual feedback, while \textit{statement-level} cohesion \cite{Krinke2007} is too fine-grained to reconstruct a programmer's train of thought.

We advocate the use of slice-based cohesion metrics on the \textit{level of variables} and showcase its use by analyzing two \textit{Java} implementations of the same example problem (implemented by an undergraduate student and by a professional programmer). We demonstrate that trains of thought can be identified in the program construction sequence. Our data sets and tools are publicly available \cite{Kesselbacher2020}.

\section{Related Work}
\label{sec:relatedWork}


Static program slicing has been introduced by Weiser~\cite{Weiser1981}, and he showed that experienced programmers tend to think in program slices when reasoning about code, for example during debugging \cite{Weiser1982}. The computation of slice-based cohesion metrics, based on slice profiles, has been introduced by Ott and Thuss \cite{Ott1993}. Their metrics (\textit{Min/Max-Coverage}, \textit{Overlap}, \textit{Tightness}, \textit{Parallelism}) make it possible to quantify the cohesion of a method (on \textit{method-level}).
These metrics have been empirically studied by Meyers and Binkley \cite{Meyers2007}, providing baseline measures, establishing that the metrics can identify deteriorating software quality as a result of changes, and investigating the inter-metric relations for sets of metrics that provide a distinct viewpoint on program code.
Krinke proposed an adaptation of the cohesion metrics on \textit{statement-level}, measuring cohesive statements to help maintainers identify parts responsible for low module cohesion that should be restructured \cite{Krinke2007}.
For programming education, static slicing has been used as an instructional method to improve students' programming skills \cite{Eranki2016} but is not widely considered.

Slice-based cohesion metrics have been adapted to formal specifications \cite{Bollin2010}, and Bollin has shown that high cohesion relates to single trains of thought, while low cohesion and small module slice intersections relate to multiple trains \cite{Bollin2013}.


LA enables researchers to analyze educational data on a completely new scale \cite{Ihantola2015, Hundhausen2017}. In programming, this includes recording program versions that are not considered finished, enabling a process-oriented analysis. Applications for text-based programming include large-scale investigations of programming mistakes in the \textit{BlueJ} programming environment \cite{Brown2018}, clustering of program versions of novice programmers solving a specific problem \cite{Blikstein2014}, investigations of a state model for programming behaviour \cite{Carter2015}, and learning curve analysis for programming concepts \cite{Rivers2016}.

We propose to use slice-based cohesion metrics in a LA setting for text-based programming education, following the ideas and findings of Bollin \cite{Bollin2013}, to improve individual assessment and feedback for learning programmers. To the best of our knowledge, there is no previous work that applies slice-based cohesion metrics in a LA setting.

\section{Showcase Study: Cohesion on Variable-Level}

\subsection{Participants and Example Problem}

\begin{figure*}
    \centering
    
    \subfloat[(a)][Example problem]{\small
\begin{tabular}[b]{rr}
\texttt{1} & \multicolumn{1}{|l}{\texttt{public class}} \\
& \multicolumn{1}{|l}{\texttt{ConvertBinary $\{$}} \\
\texttt{2} & \multicolumn{1}{|l}{\texttt{\color{white}\color{black} // Converts the input}} \\
\texttt{3} & \multicolumn{1}{|l}{\texttt{\color{white}\color{black} // array of 8 bits, and}} \\
\texttt{4} & \multicolumn{1}{|l}{\texttt{\color{white}\color{black} // prints decimal}} \\
\texttt{5} & \multicolumn{1}{|l}{\texttt{\color{white}\color{black} // and hexadecimal.}} \\
\texttt{6} & \multicolumn{1}{|l}{\texttt{\color{white}\color{black} // Example input:}} \\
\texttt{7} & \multicolumn{1}{|l}{\texttt{\color{white}\color{black} // convertBinArr(}} \\
\texttt{7} & \multicolumn{1}{|l}{\texttt{\color{white}\color{black} //  $\{$1,1,0,0,0,0,1,1$\}$)}} \\
\texttt{8} & \multicolumn{1}{|l}{\texttt{\color{white}\color{black} // Example output:}} \\
\texttt{9} & \multicolumn{1}{|l}{\texttt{\color{white}\color{black} // 195 / C3}} \\
\texttt{10} &  \multicolumn{1}{|l}{\texttt{\color{white}i\color{black} void convertBinArr(}} \\
 &  \multicolumn{1}{|l}{\texttt{\color{white}ii\color{black} int[] bN)$\{\}\}$}} \\
\end{tabular}%
}
    \subfloat[(b)][Method cohesion matrix (Student)]{
\small
\begin{tabular}[b]{r|r}
\texttt{1} & \multicolumn{1}{l}{\texttt{void convertBinArr(}} \\
 & \multicolumn{1}{l}{\texttt{\color{white}ii\color{black}int[] bN)$\{$}} \\
\texttt{2} & \multicolumn{1}{l}{\texttt{\color{white}i\color{black} double dec=0;}} \\
\texttt{3} & \multicolumn{1}{l}{\texttt{\color{white}i\color{black} String hex="";}} \\
\texttt{4} & \multicolumn{1}{l}{\texttt{\color{white}i\color{black} double hD1=0;}} \\
\texttt{5} & \multicolumn{1}{l}{\texttt{\color{white}i\color{black} double hD2=0;}} \\
\texttt{6} & \multicolumn{1}{l}{\texttt{\color{white}i\color{black} for(int i=0;}} \\
 & \multicolumn{1}{l}{\texttt{\color{white}iii\color{black} i$<$bN.length;i++)$\{$}} \\
\texttt{7} & \multicolumn{1}{l}{\texttt{\color{white}ii\color{black} if(bN[i]$==$1)$\{$}} \\
\texttt{8} & \multicolumn{1}{l}{\texttt{\color{white}iii\color{black} dec+=Math.pow(2,i);}} \\
\texttt{9} & \multicolumn{1}{l}{\texttt{\color{white}iii\color{black} if(i$<=$3)}} \\
\texttt{10} & \multicolumn{1}{l}{\texttt{\color{white}iiii\color{black} hD2+=Math.pow(2,i);}} \\
\texttt{11} & \multicolumn{1}{l}{\texttt{\color{white}iii\color{black} if(i$>$3)}} \\
\texttt{12} & \multicolumn{1}{l}{\texttt{\color{white}iiii\color{black} hD1+=Math.pow(2,i-4);$\}$$\}$}} \\
\texttt{13} & \multicolumn{1}{l}{\texttt{\color{white}i\color{black} hex=hexString(hD1)}} \\
 & \multicolumn{1}{l}{\texttt{\color{white}ii\color{black} +hexString(hD2);}} \\
\texttt{14} & \multicolumn{1}{l}{\texttt{\color{white}i\color{black} out.println(dec);}} \\
\texttt{15} & \multicolumn{1}{l}{\texttt{\color{white}i\color{black} out.println(hex);}$\}$} \\
\end{tabular}%

}
    \subfloat[(c)][Method cohesion matrix (Professional)]{\small
\begin{tabular}[b]{rr}
\texttt{1} & \multicolumn{1}{|l}{\texttt{void convertBinArr(}} \\
& \multicolumn{1}{|l}{\color{white}iii\color{black}\texttt{int[] bN)$\{$}} \\
\texttt{2} & \multicolumn{1}{|l}{\texttt{\color{white}i\color{black} int sum=0;}} \\
\texttt{3} & \multicolumn{1}{|l}{\texttt{\color{white}i\color{black} int mul=1;}} \\
\texttt{4} & \multicolumn{1}{|l}{\texttt{\color{white}i\color{black} for(int i=0;}} \\
& \multicolumn{1}{|l}{\texttt{\color{white}iii\color{black} i$<$bN.length;i++)$\{$}} \\
\texttt{5} & \multicolumn{1}{|l}{\texttt{\color{white}ii\color{black} sum+=mul*bN[i];}} \\
\texttt{6} & \multicolumn{1}{|l}{\texttt{\color{white}ii\color{black} mul*=2;  $\}$}} \\
\texttt{7} & \multicolumn{1}{|l}{\texttt{\color{white}i\color{black} out.println(sum);}} \\
\texttt{8} & \multicolumn{1}{|l}{\texttt{\color{white}i\color{black} out.println(}} \\
 & \multicolumn{1}{|l}{\texttt{\color{white}ii\color{black} convertIntToStr(sum/16)}} \\
 &  \multicolumn{1}{|l}{\texttt{\color{white}ii\color{black} +convertIntToStr(sum\%16)}} \\
 &  \multicolumn{1}{|l}{\texttt{\color{white}ii\color{black} );$\}$}} \\
\end{tabular}%
}
    
    \subfloat[(d)][Evolution of Coverage for selected variables (Student)]{\includegraphics[width=0.5\textwidth, page=4]{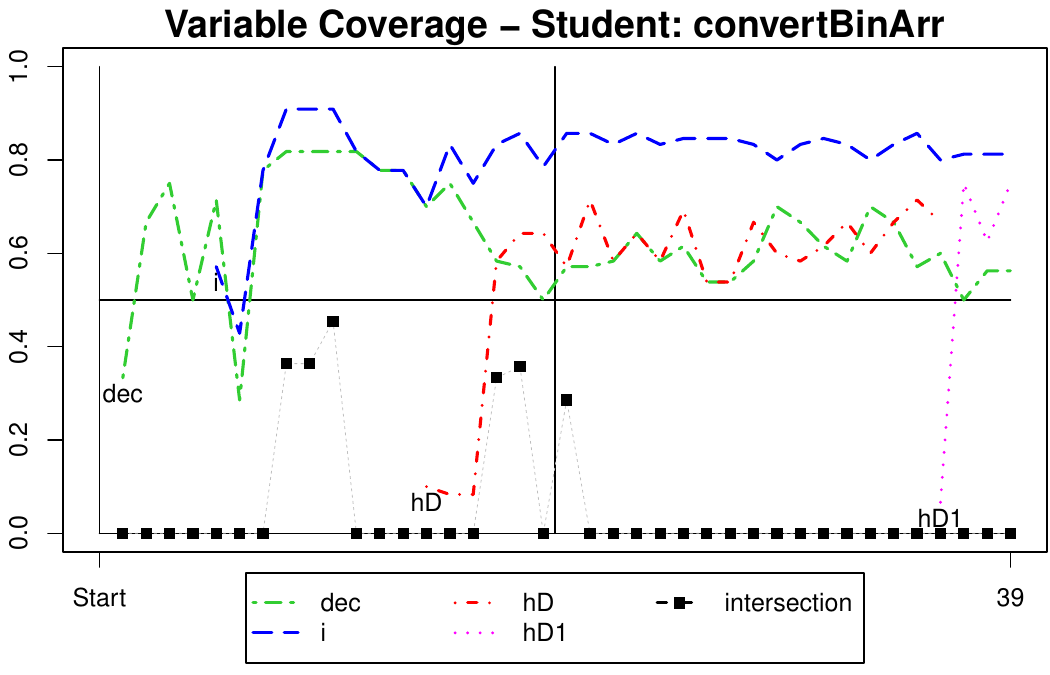}}
    \subfloat[(e)][Evolution of Coverage for selected variables (Professional)]{\includegraphics[width=0.5\textwidth, page=2]{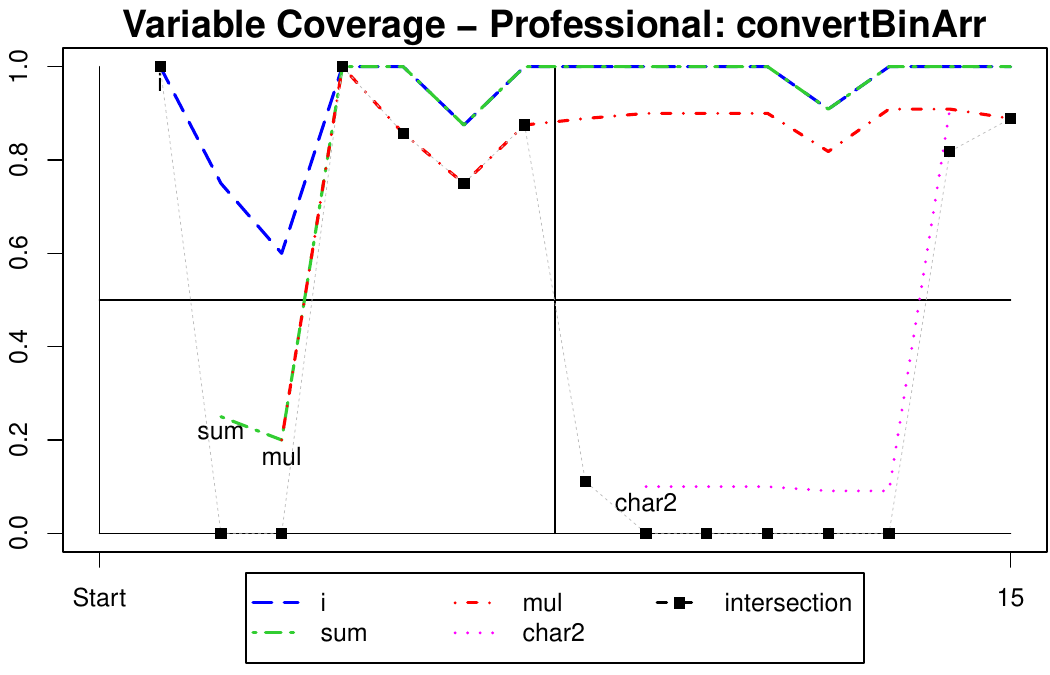}}

    \caption{The top row shows the example problem (\texttt{a}) and end versions implemented by an undergraduate student (\texttt{b}) and a professional programmer (\texttt{c}). The bottom row shows the evolution of \textit{Coverage} for selected variables and the relative method slice intersection for all consecutive, compilable program versions (undergraduate (\texttt{d}), professional (\texttt{e})). Each point on the x-axis represents a compilable program version during program construction, starting with the first compilable program change (e.g. variable declaration) and ending with the program versions shown in the top row. The final number on the x-axis represents the number of consecutive, compilable program versions. The y-axis represents the \textit{Coverage} and method intersection values.}
    \label{fig:sliceEval}
\end{figure*}

The example problem used for this showcase study is given in Figure~\ref{fig:sliceEval} (\texttt{a}). We recruited participants from two populations ($17$ undergraduate and graduate students from a Software Engineering project management course at the University of Klagenfurt, and $9$ professional programmers from an Austrian software development company) to implement the example problem in \textit{Java}. We use this example problem specifically because there are multiple ways to solve the problem: syntactically (loops and conditional branching may be used but are not required) and semantically (computing the output numbers directly from the array or computing one number from the binary array and the second one from the first number).

With the example problem's openness, it is possible to test participants of varying programming skills while also providing potential differences to be assessed with a variety of metrics. In our showcase study, we use slice-based cohesion metrics to differentiate ways to solve the example problem.

We selected two implementations to showcase the use of slice-based cohesion metrics on \textit{variable-level} in a LA setting: one implementation by an undergraduate programmer showing two trains of thought with little cohesion between parts (Section~\ref{sec:showcaseUndergrad}), and one implementation by a professional programmer that features a single train of thought and high cohesion (Section~\ref{sec:showcaseProfessional}). Both implementations are fully functional and correctly implemented. We do not claim experimental validity by the selection of the implementations.

\subsection{Collection and Computation of Cohesion Metrics}
\label{sec:metricsTools}

The implementation is recorded with an IDE-based LA approach \cite{Hundhausen2017}, recording changes on \textit{keystroke} granularity \cite{Ihantola2015} (each keystroke and the resulting program version). The data collection is part of an ongoing project with a baseline collection of \textit{keystroke} granularity \cite{Kesselbacher2020}. We convert the data into a sequence of consecutive, compilable program versions.

The participants used the IDE \textit{IntelliJ} to implement the example problem. The implementation was recorded with an IDE plugin. The consecutive program versions are stored on a headless data collection endpoint. Access to the data is possible at our data collection repository \cite{Kesselbacher2020}. We adapted the slice-based cohesion metrics of Ott and Thuss \cite{Ott1993} to compute the \textit{Coverage} for each local variable and parameter of a method, and implemented an automatic computation of slice profiles and slice-based cohesion metrics for variables as an adaptation to the Java slicer \textit{JRazor} \cite{Rama2013}. The IDE plugin, the data collection server and the Java slicer are developed and hosted at our department and available as open source\footnote{Source code available at: https://gitlab-iid.aau.at/seqtrex}.

The metrics computation is based on unions of forward and backward slices, originally called \textit{metric slices} \cite{Ott1993}. We are focusing on semantics, simply calling them \textit{union slices}.
Our computation of \textit{union slices} is similar to those of \textit{metric slices}: the backward slice is computed from the last reference of the slicing variable, the forward slice is computed from the definitions of the slicing variable that are included in the backward slice. However, we only compute the forward slice from the first definition. This way, multiple \textit{union slices} can be computed for a slicing variable: iteratively computing pairs of backward and forward slices for references and definitions of the slicing variable not covered so far, until all of them are included in at least one \textit{union slice}. Different cohesive parts of a method can be found, each captured by a \textit{union slice}. Our adaptation also makes the computation of the pairwise cohesion of all variables possible, which is not showcased here\footnote{Additional information can be found in the Ph.D. thesis of the first author, currently as work in progress: \url{https://www.aau.at/wp-content/uploads/2021/03/Kesselbacher_Thesis.pdf}}. \textit{Coverage} is computed as follows:

\begin{definition}{Coverage($M$,$v$).}
Coverage, for method $M$ and slice profile $SP$ for variable $v$, is the average of the variable \textit{union slice} lengths divided by the method length.

\[ Coverage(M,v) = \frac{1}{\vert SP_{v} \vert} \sum_{i=1}^{\vert SP_{v} \vert}{\frac{\vert UnionSlice(v)_{i} \vert}{\vert M \vert}} \]
\end{definition}




\subsection{Showcase I: Undergraduate Programmer}
\label{sec:showcaseUndergrad}

The undergraduate programmer implements two computational parts to solve the example program (Figure~\ref{fig:sliceEval} (\texttt{b}), lines \texttt{6-12}). They compute the decimal sum (\textit{dec}) and the two hexadecimal digits (\textit{hD1 / hD2}) separately, controlled by the loop variable \textit{i}. After the loop, they invoke \texttt{Integer.toHexString(...)} (line \texttt{13}, in pseudo-code) to compute the hexadecimal output, and print both results.

The separated method concerns can be observed in the evolution of \textit{Coverage} on variable-level (Figure~\ref{fig:sliceEval} (\texttt{d})). With the implementation of the second computation part (the addition of the variable \textit{hD}), the coverage of the control variable \textit{i} remains high while the coverage of the data variables \textit{dec} and \textit{hD} are only moderately high. Towards the end, the undergraduate replaces \textit{hD} with \textit{hD1/hD2}, which further decreases the coverage of \textit{dec}. Both data variables have a coverage greater than $0.5$, but the relative method slice intersection is low during the whole method construction and even $0$ for most program versions. This means that, for most program versions, there is no method statement that is included in all \textit{union slices}. Even when removing line \texttt{3} (which causes the empty intersection), the relative method slice intersection is only $0.27$.

\subsection{Showcase II: Professional Programmer}
\label{sec:showcaseProfessional}

The professional programmer follows a different approach to solve the example program. They only compute the decimal sum (\textit{sum}) in their loop (Figure~\ref{fig:sliceEval} (\texttt{c}), lines \texttt{4-6}). To handle the hexadecimal output, they implement the method \texttt{intToStr} (not shown for brevity) to compute the hexadecimal digit corresponding to the integer parameter, and feed both half-bytes to this method (line \texttt{8}).

The evolution of \textit{Coverage} on variable-level (Figure~\ref{fig:sliceEval} (\texttt{e})) shows that the professional programmer eventually achieves a high \textit{Coverage} for all variables, in contrast to the undergraduate's program construction sequence. During program construction, the relative method slice intersection decreases whenever a new variable is introduced (\textit{sum / mul / char2}). Subsequently, they are integrated or refactored (in the case of \textit{char2}, which was a temporary variable to hold half of the hexadecimal result), which results in a high coverage for all variables as well as a high relative method slice intersection. The relative method slice intersection of the end version is $\frac{8}{9} = 0.\dot{8}$ - all method nodes other than the definition of \textit{sum} in line \texttt{2} is included in every union slice. Note that the loop corresponds to three nodes in the method PDG.


\section{Discussion}
\label{sec:discussion}

\subsection{Identifying Trains of Thought in the Implementations}

Bollin argues that single trains of thought translate to slice intersections covering all predicates of cohesive \texttt{Z} specifications. Different thoughts result in smaller intersections \cite{Bollin2013}. By analogy, single trains of thought should result in cohesive method implementations and high relative method slice intersections. In this paper, we propose the notion of a \textit{cohesive} method having a relative method slice intersection $>0.5$, with the rationale that, above this threshold, all variable union slices should have method statements in common.

Following this notion, we can establish that the undergraduate implements multiple trains of thought: the variable-level \textit{Coverage} metrics show a reasonably high cohesion greater than $0.5$, the slice intersection never exceeds $0.5$ and shows that there are no cohesive method parts across all variables. It is therefore important to look at both variable-level \textit{Coverage} as well as method slice intersection measures.

The professional programmer integrates all variables into a cohesive method implementation, following a single train of thought. The slice intersection exceeds $0.5$ multiple times during the program construction, including most method parts in all \textit{union slices}. The variable-level \textit{Coverage} metrics also indicate high method cohesion.

\subsection{Limits of Cohesion Metrics and Threats to Validity}

Applying the slice-based cohesion metrics on variable-level revealed some limits. For small methods, typically used in the training of novice programmers, implementations will often result in a single union slice per variable that covers the better part of the method. 
Moreover, primitive input parameters are usually not re-defined, which makes iteratively computed \textit{union slices} less important for those parameters. For non-primitive input parameters, the approach is still worthwhile.

Moreover, threats to validity need to be addressed. First, the uncontrolled openness of the example problem can be seen as a threat to validity. There are various ways to implement a solution, which are not inherently tied to the level of programming skills. However, for this pre-study, this uncontrolled openness is a strength to collect different implementations.

Second, the recruitment of participants was uncontrolled regarding their programming skills. Undergraduate and graduate students of the curricula of \textit{Applied Computer Science} and \textit{Management of Information Systems} have participated in the study. This leads to uncontrolled variance in the students' programming experience. The professional programmers reported a mean programming experience of $17.29\pm9.62$ years. We mitigate this threat by only reporting a selected case study.

A third threat to validity arises from the adaptation to the \textit{Java} slicer \textit{JRazor}~\cite{Rama2013} to compute the slice-based cohesion metrics, which was not validated elsewhere. To mitigate this threat, we double-checked the reported metric results by hand.

\subsection{Use of the Cohesion Metrics for Education and Training}

The outlined IDE-based LA approach facilitates real-time assessment and feedback of program construction sequences. We see different use cases to employ the reported cohesion metrics and visualizations in education and training. First is the direct use by educators, which provides an additional, automatically processed, aspect during the assessment of student-written programs and supports the educators to give in-depth feedback regarding the program construction as a process. This is especially valuable when assessing long methods.

Second is the use of IDE-based \textit{interventions} \cite{Hundhausen2017} to augment the information accessible to students during programming, including \textit{variable-level} cohesion reports on demand, or cohesion information in the programming view for each variable.

By incorporating the topic of cohesion in the process of learning to program, students benefit from experiencing program structures that lead to low and high cohesion, respectively. Programming tasks with target cohesion values can be designed so that students practice the creation of highly cohesive programs - starting from program code that is functional but non-cohesive. Such tasks teach students the design principle of \textit{single responsibilities} in a quantifiable manner. Students are thereby empowered to develop a holistic view of programming: not writing single source code statements, but constructing semantically meaningful program parts.

In light of the neo-Piagetian learning process model established by Lister, we suggest that students need to operate on the third, \textit{concrete operational}, stage (being capable of '\textit{a purposeful approach to writing code}' \cite{Lister2016}) in order to benefit most from the cohesion information. However, they can thereby be supported in their process of developing expert programming skills towards the last, \textit{formal operational}, stage.

\section{Conclusion and Future Work}
\label{sec:conclusion}

To better and faster prepare our students for future challenges, we aim to improve the individual assessment and feedback of programming skills by incorporating slice-based cohesion metrics into IDE-based LA research. In this contribution, we report on a pre-study to showcase the use of the slice-based cohesion metric \textit{Coverage} on \textit{variable-level}.

We showcase the metric, investigating two \textit{Java} implementations of an example problem (conversion of a binary input array to decimal and hexadecimal numbers), done by an undergraduate and a professional programmer. Different trains of thought during program construction can be identified by measuring the intersection of all variable union slices. Our data sets and tools can be openly accessed (see Section~\ref{sec:metricsTools}).

We plan to conduct a systematic evaluation of the relation between programming skills and slice-based cohesion metrics in the future. A tabulated experimental design can be formed by controlling the programming skills to form homogeneous experimental cohorts, while separating implementations by strategic implementation approaches.

\section{Author Profile}

\textbf{Max Kesselbacher} is a University Assistant at the Department of Informatics Didactics at the University of Klagenfurt. He completed graduate studies in computer science as a teaching profession as well as computer science with a focus on software engineering and now aims to improve the education of computer science and software engineering in particular. He contributes to informatics workshops, organizes coding contests, and operates a platform to investigate the lasting effects of interventions. In his Ph.D. thesis he focuses on improving the acquisition of programming skills by employing a learning analytics approach, measuring program construction, and making individual support and feedback possible.
 
\textbf{Andreas Bollin} is a Full Professor at the University of Klagenfurt and head of the Department of Informatics Didactics. He worked on numerous projects dealing with computer science education and different types of new media in education. He authored and co-authored over 70 international peer-reviewed publications, combining informatics didactics, serious games, programming, formal methods, and software engineering. He heads a non-profit center for teaching computing science to the public (the ''Informatikwerkstatt'') with more than 14.000 attendees in the past four years. The focus of his research includes computing education in primary and secondary education, educational and serious games, computational thinking, competency, maturity models and gender/personality aspects in teaching.

\bibliographystyle{IEEEtran}
\bibliography{seeng21-slicing-paper}

\end{document}